# Overflow of a dipolar exciton trap at high magnetic fields

S. Dietl[1,2], K. Kowalik-Seidl[3], D. Schuh[4], W. Wegscheider[5], A. W. Holleitner[1,2*], and U. Wurstbauer[1,2*]

[1]Walter Schottky Institut and Physics Department, Am Coulombwall 4a, Technical University Munich, D-85748 Garching, Germany

[2]Nanosystems Initiative Munich (NIM), Schellingstr. 4, 80799 München, Germany

[3]Center for Nanoscience and Fakultät für Physik, Ludwig-Maximilians-Universität, Geschwister-Scholl-Platz 1, 80539 München, Germany

[4]Institute of Experimental and Applied Physics, University of Regensburg, D-93040 Regensburg, Germany

[5]Solid State Physics Laboratory, ETH Zurich, 8093 Zurich, Switzerland

Abstract

We study laterally trapped dipolar exciton ensembles in coupled GaAs quantum wells at high magnetic fields in the Faraday configuration. In photoluminescence experiments, we identify three magnetic field regimes. At low fields, the exciton density is increased by a reduced charge carrier escape from the trap, and additionally, the excitons' emission energy is corrected by a positive diamagnetic shift. At intermediate fields, magnetic field dependent correction terms apply which follow the characteristics of a neutral magnetoexciton. Due to a combined effect of an increasing binding energy and lifetime, the exciton density is roughly doubled from zero to about seven Tesla. At the latter high field value, the charge carriers occupy only the lowest Landau level. In this situation, the exciton trap can overflow independently from the electrostatic depth of the trapping potential, and the energy shift of the excitons caused by the so-called quantum confined Stark effect is effectively compensated. Instead, the exciton energetics seem to be driven by the magnetic field dependent renormalization of the many-body interaction terms. In this regime, the impact of parasitic in-plane fields at the edge of trapping potential is eliminated.

*corresponding authors: holleitner@wsi.tum.de and wurstbauer@wsi.tum.de

The investigation of many-body effects and quantum phase transitions in dense ensembles of dipolar excitons has motivated different approaches to increase their density[1]. In particular, the exciton density can be increased by elongating the radiative lifetime of the excitons, which is typically achieved by reducing the spatial overlap of the electron and hole wave functions. This has been realized in tunnel-coupled quantum well structures (CQW), where photogenerated electron-hole (*e-h*) pairs are spatially separated by the application of a perpendicular electric field. Hereby, the photogenerated electrons tunnel to one quantum well and the holes to the other one, but they are still bound to each other by the Coulomb interaction. Such indirect or dipolar excitons (IX) have a permanent dipole moment and they feature long lifetimes up to several microseconds[2]. Huge efforts have been made over the last years to further increase the IX density as required for quantum phase transitions to achieve exciton fluids[1] and even a Bose-Einstein condensation (BEC)[1]. Promising approaches to increase the IX density include a lateral confinement of the IX ensembles, e.g. by strain in the hosting materials[3–6], by shaping the excitation laser light profile[7] and most commonly, by tailored electric field landscapes and gradients within the plane of the quantum well structures[8–16]. Hereby, electrostatic traps for the IXs can be formed, and it has been demonstrated that one can create dense ensembles therein with the temperature of the IXs close to the lattice temperature[17,18]. The functionality of such electrostatic traps is typically altered by unwanted parasitic effects[12]. These effects comprise in-plane electric fields, which can lead to the dissociation of the IXs, and consequently, charge carrier escape processes, which give rise to the excess of one sort of charge carriers within the traps, i.e. photogenerated electrons or holes[19]. The presence of such unbound excess carriers in the trap can weaken the Coulomb interaction between the bound *e-h* pairs by screening and scattering effects [20–23]. Moreover, charged excitons, so-called trions, can form with an emission energy lower than the one of the IXs[24,25]. Such trions obey Fermi-Dirac statistics rather than Bose-Einstein statistics and are therefore not suitable to observe a BEC[26]. An experimental way to reveal the presence and the density of excess charge carriers is given by the application of a magnetic field perpendicular to the quantum well plane[27]. The application of large magnetic fields can greatly suppress the formation of charged excitons. The underlying reasons are localization processes, which prevent the free charge carriers from escaping the electrostatic trap region[27]. As a direct consequence, the IX density can be enhanced by the localization of both electrons and holes within the trap.

Generally speaking, the application of magnetic fields significantly modifies the exciton wave functions in the CQWs and turns the excitons into so-called magnetoexcitons. In particular, the exciton binding energy is enlarged in the presence of strong magnetic fields due to shrinkage of the exciton wave function[28]. The exciton oscillator strength is altered by the magnetic field, and the application of a field parallel to the CQWs affects the excitons' dispersion[29,30]. Magnetoexcitons have been investigated theoretically[31–35], and a variety of fascinating effects including quantum phase transitions to an exciton liquid or BEC have been suggested[32,35–39]. Indirect magnetoexcitons built from indirect, dipolar excitons feature an increased effective mass reducing the IX localization[29]. As a direct consequence, the lifetime of indirect magnetoexcitons is increased compared to IXs at zero magnetic field. In principle, the application of an external magnetic field normal to the QW plane results in an opposite Lorentz force experienced by the electron and hole. The corresponding relative displacement results in an additional dipolar moment[30]. The application of perpendicular electric fields pushes the electrons and holes away from each other[28]. The application of an electric field perpendicular to the plane of the CQWs induces a cross-over from a direct to an indirect exciton ground state. For extended exciton ensembles, this cross-over shifts to larger electric field values under the application of magnetic fields as was shown experimentally[40] and theoretically[28].

In this manuscript, we study the impact of large magnetic and electric fields on laterally confined dipolar excitons that are photogenerated in CQWs. We demonstrate that a trap of IXs can 'overflow' at high magnetic fields. The excitons are electrostatically trapped in micrometer scale lateral traps to realize dense ensembles of IXs at low temperatures. In magnetic field dependent photoluminescence experiments, two critical magnetic field values are found, which we baptize $B^*$ and $B^{**}$. In the low field regime ($B < B^*$), the exciton ensembles occupy the trapped region in co-existence with free charge carriers[27], and the IX energy is dominated by dipole-dipole interactions. This low magnetic field regime has been investigated in great detail in an earlier publication[27]. For intermediate magnetic field values ($B^* < B < B^{**}$), the ground state energy of the IXs is dominated by Landau Level (LL) formation of the electron and hole bands. In principle, both are single particle energies. However, the exciton emission energy is shifted to higher values by the increased binding energy as well as density-dependent dipole-dipole interactions. In this magnetic field regime, we demonstrate that the exciton density within the traps is continuously increased with an increasing magnetic field. We interpret this situation such that all excess carriers are localized within the trap and that the increasing binding energy allows for a larger exciton density[27]. At the second critical field value $B^{**}$, we observe that the IX emission energy can be independent from the electrostatic trapping potential. As a function of laser intensity, the traps can overflow. We attribute this unusual behavior to the situation that the trap is completely filled with neutral indirect magnetoexcitons and that the many-body interaction induced renormalization of the exciton energy compensates the electrostatic trapping potential. As a consequence, the excitons are no longer effectively confined by the trap potential. They are able to diffuse out of the excitation spot.

This diffusion effectively lowers the exciton density and hence their many-body energy inside the trap. Moreover, we observe another impact of the trapping potential, namely that the electric field induced shift of the cross-over from direct to indirect exciton ground state remains unaffected by the external magnetic field. This finding is different to the observations by Butov *et al.*[40] and Morales *at al.*[28] for non-trapped exciton systems at high magnetic fields. We attribute the different behavior between laterally confined (trapped) and extended exciton systems also to density-dependent renormalization effects of the IX energies.

The exciton ensembles are photogenerated in GaAs-based CQWs with AlGaAs-barriers grown by molecular beam epitaxy (MBE) on semi-insulating (001) GaAs substrates. The heterostructure consists of two 8 nm-wide GaAs quantum wells separated by a 4 nm-thick $Al_{0.3}Ga_{0.7}As$ tunnel barrier. The valence and conduction band structure is tilted under the application of an electric field perpendicular to the QW plane [Fig. 1(a)]. The CQWs are embedded in a 370 nm-thick field-effect device with an MBE grown heavily n-doped GaAs back-gate electrode and 5 nm-thick semi-transparent top-gate electrodes made from titanium by e-beam evaporation. The application of an electric field shifts the energy of the IXs with respect to the energy of the direct excitons (DX) by the so-called linear quantum confined Stark (QCSE) effect by an amount of $-\vec{p} \cdot \vec{F}$, where $\vec{p}$ denotes the effective exciton dipole moment and $\vec{F}$ the electric field. The electric field can be approximated as $\vec{F}$ = - ($V_G$ + $V_{Schottky}$) / $s$, with $V_G$ the voltage between a top gate and the global bottom gate, and $V_{Schottky}$ = -0.7 V the effective barrier height of the metal Schottky barrier. The parameter $s$ = 370 nm describes the vertical distance between the top and bottom gates. Fig. 1(b) sketches the cross-section and lateral geometry of the investigated devices. A circular center top gate ($V_C$) is surrounded by a second top gate called guard gate ($V_G$). Both top gates are biased with respect to the back gate. This allows for a precise tuning of the lateral potential landscape for the IXs [8,9]. We show results from two different devices F1 and F2 with different trap diameters prepared from the same wafer. The trap diameter of device F1 (F2) is 23 µm (16 µm).

All measurements are carried out at a bath temperature of $T_{bath}$ = 4.2 K in an optical cryostat equipped with a superconducting electromagnet up to 9 T in Faraday configuration. We utilize a continuous wave laser diode with an energy of $\hbar\omega_{laser}$ = 1.823 eV to excite the CQWs. The laser light is focused on the sample with an aspherical lens with a 2 mm focal length. The spot diameter is approx. 10 µm for the excitation and approx. 1 µm for the detection of the exciton emission. In the experiments, the maximum laser intensity within the excitation spot is $P_0 \approx$ 50 µW (64 W/cm$^2$), and is adjusted by neutral density filters as discussed below. By the mentioned experimental conditions, we generate IX ensembles with variable densities as high as $n_{IX}$ ~ $10^{11}$ cm$^{-2}$ under the center gate[27]. For $V_C < V_G$, the potential landscape is such that electrons can effectively escape from the trap region at zero magnetic field, while the less mobile holes are confined under the center gate[19]. The IX density $n_{IX}$ can be deduced from the density-dependent blue-shift of the IX energy in PL experiments[41]. The resulting two-dimensional excess holes in the trap sum up to a hole density $n_h$ of the same order of magnitude as those of the IX ensemble[27]. Figure 1(c) shows typical photoluminescence spectra from the IXs at zero magnetic field for device F1. The spot position is located at the center gate electrode. For $V_G = 0$ V and $V_C = -0.1$ V, the device acts as an exciton trap (black line) and for gate parameters $V_G = 0$ V and $V_C = 0.1$ V, it acts as an anti-trap (grey line). In the case of the trap situation, the IX energy is red-shifted and the emission intensity is reduced with respect to the anti-trap situation. Generally, a larger electric field results in an enlarged separation of electron and holes and therefore, in a reduced overlap of electron and holes wave functions in the vertical direction. As a consequence the IX exhibit longer lifetimes for increasingly negative gate voltages[42,43].

Figure 1(d) depicts the evolution of PL energy and intensity of laterally confined DXs and IXs in the presence of a magnetic field normal to the QW plane. Under the excitation spot on the center gate of device F1, the excitons experience a trap potential with parameters $V_G = 0$ V and $V_C = -0.4$ V. At low magnetic fields, the emission energy of the excitons increases quadratically with the magnetic field. This behavior can be understood by the so-called diamagnetic shift given by $\Delta E_{dia} = (e^2 a_B^2 / 8\mu) B^2$, with $e$ the elementary charge, $a_B$ the exciton Bohr-radius, and $\mu$ the reduced in-plane mass of the exciton[44]. The relationship holds as long as the energy shift is small compared to the exciton binding energy[44–47]. In the device under investigation, the diamagnetic shift of the IX ensembles can be determined to be $\Delta E_{dia} \approx 0.06$ meV/T$^2$. We observe an additional blue-shift in the PL experiments that we assign to an increase of the IX density $n_{IX}$ with magnetic field, which we interpret in terms of carrier localization effects. In the trap configuration, electrons can escape from the center gate region leaving excess holes behind. By applying a magnetic field normal to the CQWs, the Lorentz force constrains the electrons and holes onto orbital trajectories. Hereby, also the electrons are effectively localized under the center electrode[27]. As a direct consequence, the IX density is increased, and simultaneously, the amount of excess holes is reduced. We interpret the transition to a new regime at $B^* \approx 1.5$ T [Fig. 1(d)], such that all electrons are localized in the trap region under the central gate[27].

Depending on the relation between the distance between the QW centers $d \approx 12$ nm and the magnetic length $l_B = (\hbar/eB)^{1/2}$, different ground states of the electrons and holes as well as excitons are predicted. The interlayer electron-hole interaction is given by the Coulomb interaction. This attractive interaction is expected to dominate at small magnetic field values (i.e. $d/l_B < 1$), whereas the repulsive electron-electron and hole-hole interactions given by $l_B$ are expected to dominate at large magnetic field values (i.e. $d/l_B > 1$)[45]. Many-body phenomena such as the exciton binding energy and exciton-exciton interactions are neglected in this first approximation. For $B > B^*$, we observe a somewhat linear increase of the PL emission energy with magnetic field. We attribute this dependence to the quantizing effect of a magnetic field on the single particle levels of conduction and valence bands. In other words, the photogenerated electrons and holes in the CQWs fill the electron and hole Landau levels (LLs). The energy of the $n^{th}$ electron and hole LLs is determined by the Landau formula $E^{e(h)}_{LL} = (n + ½) \, e \cdot B / m^*_{e(h)}$, with $m^*_{e(h)}$ the effective electron (hole) mass. The electrons and holes residing on the LLs can form bound states with magnetic field dependent binding energies. This statement holds for both the DXs and IXs. In principle, the interpretation of a LL formation explains that the emission energies for both DXs and IXs increase monotonically vs. magnetic field. The binding energies are given by $E_B^{DX} \propto 1/l_B$ and $E_B^{IX} \propto 1/(l_B^2 + d^2)^{1/2}$, respectively[45]. Hereby, the different binding energies establish different slopes of the magnetic field dependent PL energies for the DXs and the IXs in the regime $B > B^*$. This is clearly visible in the magnetic field dependent emission spectra of Fig. 1(d). There are additional contributions to the energies of the magnetoexcitons, e.g. due to magnetic field dependent effective mass enhancement of the excitons[29], to increased lifetimes[45] as well as to an altered dipole-dipole interaction. For a fixed exciton density, the latter is weakened with an applied magnetic field due to a reduced exciton radius[45,32]. This reduction of the dipole-dipole interaction lowers the exciton energy, while an increase in the exciton density at a fixed magnetic field value still results in an increase of the IX emission energy due to the repulsive dipole-dipole interaction. Furthermore, an increased effective exciton mass reduces the diffusion of the excitons out of the excitation spot which can give rise to an increase of the IX density under the excitation spot. Another source for magnetic field dependent increase of the IX density is again an increased lifetime. Moreover, the exciton ensembles are laterally confined by the trapping potential between the center and guard gate electrodes.

So, there are several competing contributions to the IX emission energy also for the trapped ensembles in magneto-PL experiments. These are the gap energy of the GaAs-based CQWs $E_G$, the energy shift due to the applied electric field in growth direction via $E_{field} = -\vec{p} \cdot \vec{F}$, the already discussed LL energy of the bound electron and hole levels $E_{LL} \propto B_\perp / \mu$, the binding energy $E_B^{IX} \propto 1/(l_B^2 + d^2)^{1/2}$ and the dipole-dipole interaction dependent blue-shift given by the IX density $\Delta E_{shift} \propto \Delta n_{IX}(B, P, V)$ that depends on the applied magnetic field, the laser excitation power and the electric fields within the trapping potential. The latter are adjusted by the applied gate voltages.

On top, we observe small discontinuities in the magnetic field dependent evolution of the emission spectra which particularly superpose a general linear dependence of the emission energy. The discontinuities are marked by asterisks in Fig. 1(d), and we attribute them to a modification of the Coulomb interaction within the electron and hole systems. Here, we follow the argument that the Coulomb interaction in the LLs strongly depends on the integer filling factors[48–52,52]. In this interpretation, the discontinuities mark jumps of the chemical potential from a higher to the next lower LL. As a matter of course, the jumps depend on the magnetic field value and on the charge carrier densities, which is consistent with a change in the filling factor $\nu = n_{e(h)}h/eB$, with $n_{e(h)} \approx n_{IX}$ the electron (hole) density in the CQWs. The former expression neglects spin degeneracy. We note that in our experiment, the spacing of the discontinuities does not exhibit a $1/B$ dependence, which is consistent with a significant variation of $n_{IX}$ for $B > B^*$.

In order to further deconvolute the different magnetic field dependent contributions to the exciton energy, we compare emission spectra taken for fixed magnetic field values of $B = 0$ T, 3 T, 7 T and 9 T on device F2 [Figs. 2(a) to 2(d)]. All spectra are taken at an identical laser power $P_0$ and a constant guard gate ($V_G = 0$ V), while the trapping potential is varied ($V_C$ is tuned within the range of -0.6 V < $V_C$ < 0.4 V). Independent from the magnetic field, the trap to anti-trap transition occurs at $V_G \approx V_C$ as marked by the vertical dashed line in Figs. 2(a) to 2(d). The corresponding IX energy is highlighted by a horizontal arrow in each of the figures. For all investigated magnetic field values, the trap to anti-trap transition is accompanied by a modification of the electric field dependent slope of the IX energies[19]. For low negative voltages -0.1 V < $V_C$ < 0 V, we observe a linear shift of the IX emission energy with gate voltage as expected from the linear QCSE. For $B = 0$ T [Fig. 2(a)] and 3 T [Fig. 2(b)], the IX emission energy follows the QCSE in the whole trap regime (-0.6 V < $V_C$ < 0V). However, at $B = 7$ T [Fig. 2(c)], a remarkably different behavior is apparent with an almost constant emission energy for the most negative gate voltages. Simultaneously, the emission intensity is significantly enhanced. This increase in intensity is also visible in Fig. 1(d) at $B = 7$ T. For larger magnetic fields, the QCSE seems to recover for a large voltage range. In particular, at $B = 9$ T [Fig. 2(d)], the emission energy is again reduced for increasing negative center gate voltages $V_C$. However, even at this high magnetic field, there are still deviations from a linear slope, as one would expect from a linear QCSE. For a better comparison, Figs. 2(e) and 2(f) compare the IX emission energy and integrated intensity for $B = 0$ T (black symbols) and $B = 7$ T (grey symbols). The expected IX emission energy at $B = 7$ T due to the QCSE is indicated as a dashed line in Fig. 2(e). Apart from the extraordinary gate voltage dependent behavior at 7 T, the IX energy is generally increased by the application of a magnetic field for $B^* < B < B^{**}$. This observation holds for all investigated center gate voltage values as demonstrated in Fig. 2(g).

Again, there is a significant change in the magnetic field dependent behavior of the IX energy at $B^{**}$ = 7 T. For the trap configuration ($V_C$ < $V_G$ = 0V), the IX energies constitute $E_{IX}$ = 1.560 eV independent from $V_C$. For larger magnetic field values ($B > B^{**}$ = 7 T) the IX emission energies decrease with increasing center gate voltage and hence electric field.

We would like to point out that the electric field dependent crossover from IX to DX ground state always occurs at $V_C \approx 0.2$ V, i.e. independent from the applied magnetic field. This independency from the applied magnetic field opposes reports in literature on magnetoexcitons without trapping landscapes[28,40]. We assign this difference in the magnetic field dependent behavior to the lateral confinement of the IX system in our study, because in a trap, denser IX ensembles can be created compared to extended geometries without lateral confinement of the exciton ensemble. A transition to a somewhat extended exciton system is achieved at $V_C \approx V_G$. At finite magnetic fields, this situation resembles an ensemble of laterally non-confined magnetoexcitons. Indeed, as plotted in Fig. 2(g) (diamonds), the emission energy for such an extended ensemble is different to the confined situation. The emission energy monotonously increases for the whole range of examined magnetic field values. Moreover, we do not observe any critical magnetic field values in the emission characteristics of such an extended exciton system at the trap to anti-trap transition in our experiments.

Generally, we interpret the magnetic field and gate voltage dependent blue-shift in the extracted exciton energies in Figs. 2(g) as an increase of the exciton density in the trap. At zero magnetic field, the blue-shift $\Delta E_{shift}(n_{IX})$ of the emission energy due to dipolar repulsion can be expressed as[41]:

$$\Delta E_{shift}(n_{IX}) = \frac{e^2 d}{\epsilon_r \epsilon_0} n_{IX} \cdot f(n_{IX}, T) ,$$

with $\epsilon_0$ the vacuum permittivity, $\epsilon_r$ = 12.9 the relative dielectric constant, and $f(n_{IX},T)$ a correction factor to $n_{IX}$ with $0.1 \leq f(n_{IX},T) \leq 0.6$ taking into account effective interaction potentials[18]. To access the exciton densities at finite magnetic fields, we determine the deviation of the exciton emission energy from the expected value from a linear slope due to the QCSE [double arrow in Fig. 2(e)]. Fig. 2(h) depicts the extracted values of this apparent additional exciton density $\Delta n_{IX}$ as a function of the magnetic field for $V_C$ = -0.6 V and $V_G$ = 0 V. The extracted values $\Delta n_{IX}$ are minor in the low field region as one expects for $B < B^*$, and $\Delta n_{IX}$ increases monotonically to more than $\Delta n_{IX}(B) > 1.6 \cdot 10^{11}$ cm$^{-2}$ for $B^* < B < B^{**}$. For $B > B^{**}$, it decreases again. It is worth mentioning that at zero magnetic field, the intensity of the photogenerated IX system is in the order of $n_{IX} \sim 10^{11}$ cm$^{-2}$ at the used power density $P_0$. Consequently, the IX density is approximately doubled at the critical magnetic field of $B^{**}$ according to this analysis. In turn, we interpret the magnetic field dependent increase in the PL energy and intensity for $B^* < B < B^{**}$ as an effective increase of the repulsive dipolar interaction in the dense exciton ensemble. Microscopically, we assign the extracted increase of the exciton density to the combined effects of an extended lifetime[45], a reduced exciton diffusion as a result of the increased effective mass[29], an increasing LL degeneracy and a presumably altered oscillator strength in absorption and emission of photons[53] with increasing magnetic field. At the critical magnetic field $B^{**}$, we assume that only the lowest LL with $N = 0$ are occupied for conduction and valence bands. This assumption is based on the fact that the expected exciton density is smaller than $n_{IX} \leq 3.3 \cdot 10^{11}$ cm$^{-2}$ and therefore the corresponding quasi-Fermi energies for electrons and holes are in the $N = 0$ LL.

With this discussion, it is again noteworthy that the exciton emission within the trap is independent from $V_C$ for $B^{**}$ (7T) [dashed line in Fig. 2(g)]. It occurs at ~1.56 eV which roughly equals the emission energy of the guard gate at the magnetic field $B^{**}$ [arrow in Fig. 2(c)]. In first approximation, the trapping potential $E_{trap}$ is the potential difference between the guard and the center gate electrodes: $E_{trap} = |V_G - V_C| \cdot ed/s$. Correspondingly, Fig. 3 sketches the potential difference for an empty trap potential at zero (black solid line) and finite magnetic field (dashed line). In our interpretation, the magnetic field dependent blue-shift $\Delta E = \Delta E_{shift}(n_{IX}[B,P,V])$ completely compensates the trapping potential at $B = B^{**}$. The main contribution to the blue-shift of the emission energy is caused by a density dependent increase in the dipolar repulsion. For $B \geq B^{**}$, $\Delta E_{shift}(n_{IX}[B,P,V])$ would exceed the emission energy below the guard gate. In turn, the excitons spill from the trap region and diffuse into the guard region. The overflow of excitons from the trap region is further enhanced by the enlarged exciton diffusivity, when electrons and holes reside in the lowest ($N = 0$) LL compared to higher LLs [54]. This scenario results in a reduced exciton density for $B > B^{**}$ and hence, in a reduced repulsive dipole-dipole interaction as well as a reduced blue-shift of the emission energy at the position of the trap. Such a self-renormalization is expected until the density is high enough to compensate the exciton diffusion out of the excitation spot to regions with a lower energy within the effective potential landscape. Our interpretation is consistent with the experimental fact that no critical fields and no overflow of excitons from the center gate region can be observed for the condition with trap and guard at identical potential (i.e. $V_C = V_G$), because in this situation, there is no lateral confinement (trapping) in the first place. We note that the effective potential landscape exhibits certain fluctuations, e.g. caused by exciton-exciton scattering, potential fluctuations within the heterostructure, and scattering on background impurities. In this respect, the width of the fuzzy line in Fig. 3 resembles the FWHM at $B^{**}$ of the IX emission energy under the realistic experimental conditions.

In order to corroborate the above interpretation, we measure the exciton PL energy as a function of the trap potential for different incident laser powers at the critical magnetic field value $B^{**} = 7$ T (Fig. 4). The spectra are taken for device F2 with a fixed guard gate voltage of $V_G = 0.1$ V. Therefore, the trap to anti-trap transition occurs for $V_C = V_G = 0.1$ V. The transition from direct to indirect ground states at $V_C \approx 0.2$ V is again unaffected from the magnetic field. The emission spectra excited with the lowest laser power $P = 0.08 \cdot P_0$ ($P_0 \approx 50$ µW) are plotted in Fig. 4(a), for an intermediate laser power of $P = 0.43 \cdot P_0$ in Fig. 4(b), and for high laser power $P = 1 \cdot P_0$ in Fig. 4(c). The spectra are taken at $B^{**} = 7$ T, and the emission energy is plotted as a function of the center gate voltage between -0.5 V < $V_C$ < 0.2 V. For the lowest excitation power [Fig. 4(a)], the blue-shift of the exciton energies follows closely a linear behavior of the QCSE as expected in the trap region[28]. The linewidth of the exciton emission is rather narrow ($FWHM \approx 2$ meV) and homogenous in the investigated potential range. For the intermediate laser power $P = 0.43 \cdot P_0$ [Fig. 4(b)], the center gate voltage dependent blue-shift of the exciton energies is reduced and the slope deviates from the linear behavior of the QCSE. Simultaneously, the linewidth changes with the potential in the investigated voltage range and it is significantly broadened ($FWHM \approx 10$ meV). This trend continuous until $P = P_0$. At the highest laser power, the exciton emission energy is again found to be rather constant in the trap regime of the device. In turn, the data presented in Fig. 4 demonstrate that the observations are independent from the trap diameter and also depth. For the highest laser power $P_0$, the emission intensity for $B = B^{**}$ is again significantly enlarged, and the linewidth is broadened. Details to the lineshape of the spectra are shown in Fig. 4(d), where individual spectra for different magnetic fields are plotted in a waterfall representation. The power dependent measurements strongly corroborate our initial interpretation of an overflow of the exciton trap at $B^{**}$. Along this line, the magnetic field induced increase of the exciton density by more than a factor of two results in an increased blue-shift $\Delta E_{shift}(n[B,P,V])$ due to repulsive dipole-dipole interaction. The additional magnetic field dependent many-body correction terms to the IX energy at $B = B^{**}$ completely compensate the trap potential $\Delta E_{shift}(n[B^{**},P,V]) = E_{trap}$ for an initially dense exciton ensemble excited at high laser power. It is remarkable that the complex interplay between additive and subtractive many-body interaction driven magnetic field dependent contributions to the IX energy result in a universal critical magnetic field $B^{**}$, where the trap can be completely filled with IXs. A requirement for the observed universal behavior is that the trap is already sufficiently filled with photogenerated IXs at $B = 0$T. This requirement has been realized by an adequate high power of the excitation laser in our experiment. The constant density in the trap is maintained by the overflow of excess excitons into the guard region. We surmise that the spilling of the excitons is promoted by that fact that the IXs reside inside and outside the trap in the lowest LL ($N = 0$), where the diffusivity is reported to be enhanced compared to higher LLs[54].

In summary, we study the magnetic field dependent emission characteristics of dense dipolar exciton ensembles realized in CQWs. The dipolar excitons are laterally confined by micrometer sized, electrostatic traps. The magnetic field is applied normal to the CQWs. We identify three different magnetic field regimes in the photoluminescence spectra of the dipolar excitons, that are separated by two critical magnetic fields $B^* = 1.5$ T and $B^{**} = 7$ T. In the low-field region ($B < B^*$), the IX energy is increased by a diamagnetic interaction term ($\Delta E_{dia}$)[44]. Furthermore, the IX density is enlarged due to the reduced escape of photogenerated electrons from the tapped region forced by magnetic field dependent carrier localization[27]. For strong magnetic field values $B > B^*$, the exciton energy is corrected by various terms, including the formation of Landau levels for the conduction and valence bands $E^{c(v)}_{LL}(B)$, a modified binding energy $\Delta E_B^{IX}(B)$, and a strongly affected dipolar repulsion depending on the exciton density $n_{IX}$. We observe an overall blue-shift $\Delta E_{shift}(n_{IX}[B,P,V])$ of the exciton emission as a function of the magnetic field. We deduce that the density of the trapped dipolar excitons roughly doubles by a value of $\Delta n_{IX}(B) > 1.5 \cdot 10^{11}$ cm$^{-2}$ in the investigated range of magnetic fields up to 7 T. In our interpretation, this increase is mainly caused by an increasing exciton binding energy and an extended lifetime with magnetic field. For already highly dense IX ensemble in the trap at zero magnetic field ($n_{IX}(0T) > 1 \cdot 10^{11}$ cm$^{-2}$), a second critical field $B^{**}$ exists. At this field value, the emission energy of the dipolar excitons is constant and does therefore not follow the expected linear slope of the quantum confined Stark effect (QCSE). For $B > B^{**}$, the IX energy is slightly decreasing. This striking universal behavior is interpreted as an overflow of the IX from the trap into the surrounding guard region. The driving force is an increase of the IX density with magnetic field that increases simultaneously the repulsive dipolar interaction between the IXs $\Delta E_{shift}(n[B^{**},P,V])$ fulfilling the condition $E_{trap} = \Delta E_{shift}(n[B^{**},P,V])$, with $E_{trap}$ the energetic depth of the trap. Generally, the overflow at $B^{**}$ is independent from the depth of the trap, as long as the laser excitation provides a sufficient number of excitons. Our findings indicate that the exciton energy is renormalized by the overflow of dipolar excitons such that their density is adjusted. It is adjusted in a way that the blue-shift due to the dipolar repulsion $\Delta E_{shift}(n[B^{**},P,V])$ compensates the depth of the trap in an universal fashion stabilizing an ensemble of neutral dipolar excitons. This ensemble seems to be robust e.g. against potential fluctuations in the heterostructure and to in-plane electric fields.

Acknowledgement

We thank X. P. Vögele, B. N. Rimpfl and G. J. Schinner for sample preparation in the clean room and L. Hammer for assistance during the measurements. This work was supported by the Center for NanoScience (CeNS), the Nanosystems Initiative Munich (NIM), LMUexcellent, the DFG Projects WU 637/4-1 and HO3324/9-1.

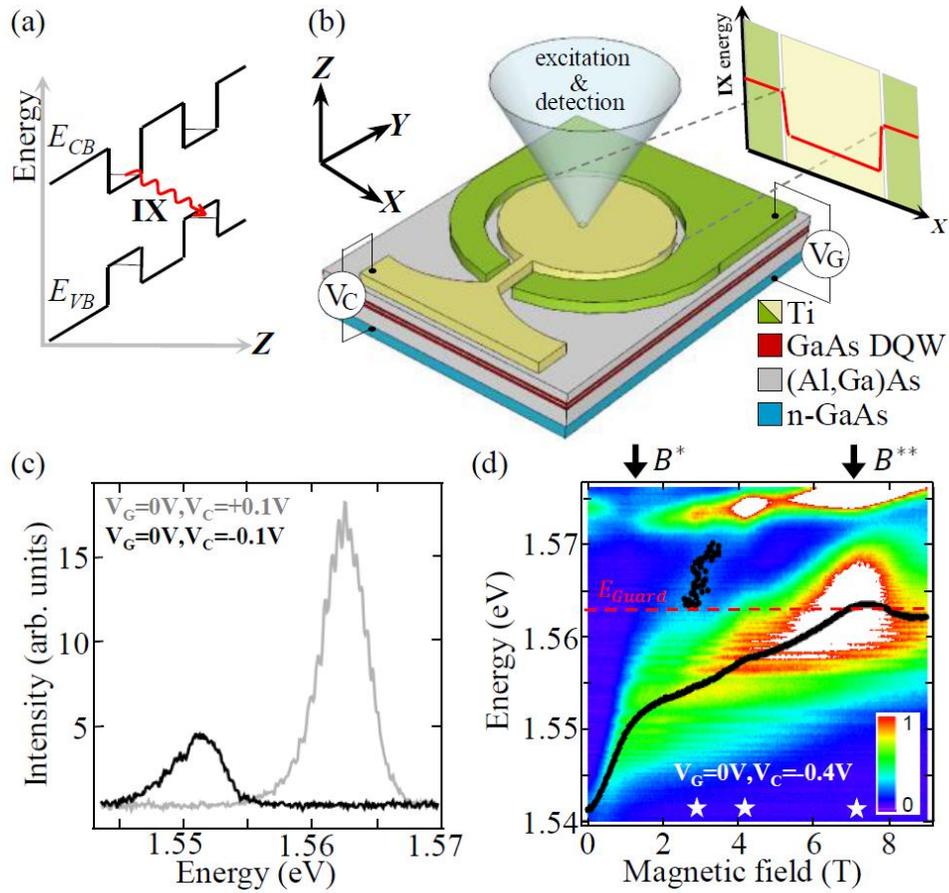

*Figure 1: (Color online) (a) Schematic band diagram of GaAs CQWs under a perpendicular electric field showing the formation of indirect exciton (IX). $E_C$ and $E_V$ denote conduction and valence band, respectively. (b) Sketch of the trapping/anti-trapping device with central gate electrode to apply $V_C$ and a surrounding guard gate electrode to apply $V_G$. Projection sketches the IX energy vs. a spatial coordinate (here, x-direction) for IXs below the gate geometry for the central gate biased to trap the IXs. (c) Photoluminescence (PL) of IXs for $V_G = 0V$ and $V_C = -0.1V$ (black line) and $V_C = +0.1V$ (grey) (device F1) (d) Magnetic field evolution of PL spectra in the trap configuration (device F1, $V_G = 0V$ and $V_C = -0.4V$). Black points and line are peak positions of the emission extracted from fits to the data. The white asterisks mark the position of kinks in the IX energy. The black arrows indicate the critical magnetic field values $B^*$ and $B^{**}$.*

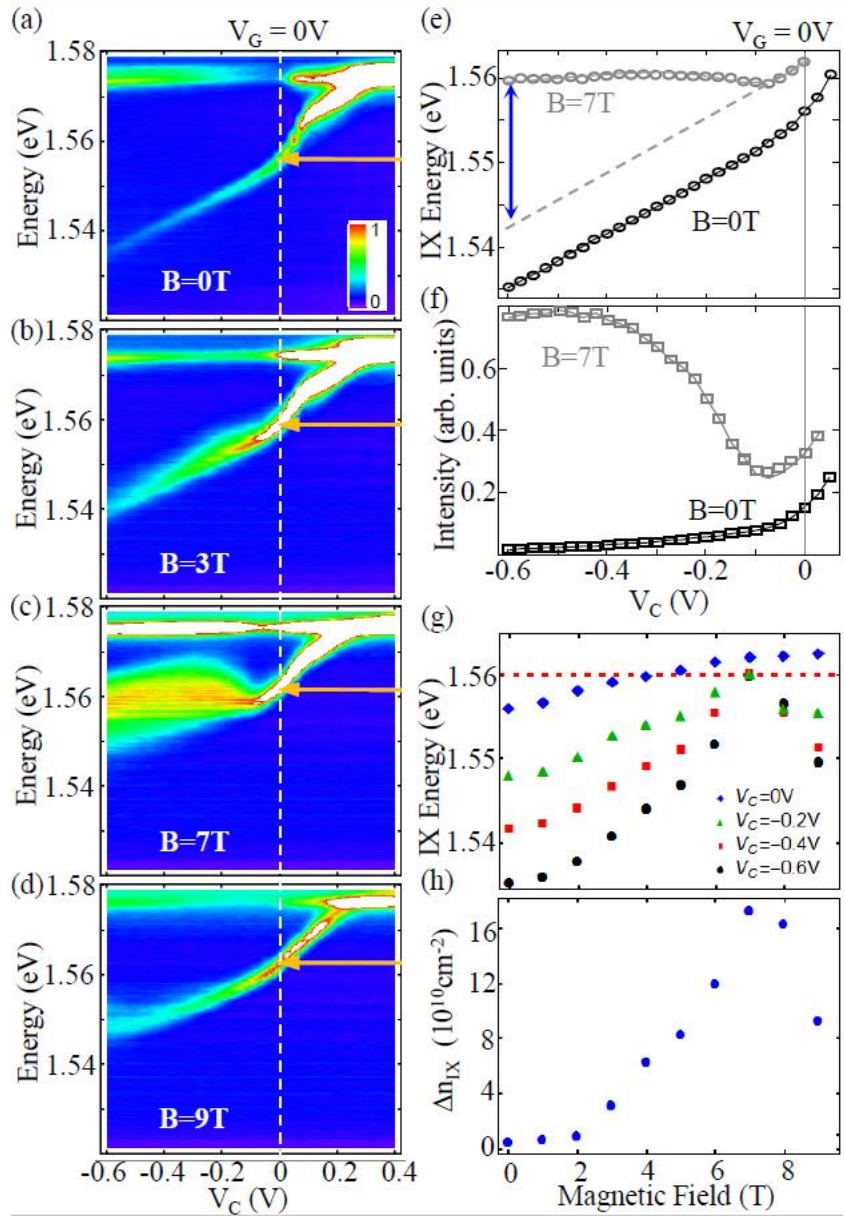

Figure 2: (Color online) (a)-(d) Photoluminescence energy of indirect excitons as function of central gate potential $V_C$ with constant guard potential $V_G = 0V$ for different magnetic field values (B = 0T, 3T, 7T, 9T from top to down). Dashed vertical lines mark the transition between trap and anti-trap potential configuration for IX (i.e. $V_C = V_G$). Arrows mark the IX energy at $V_C = V_G$. (e)-(f) Corresponding emission energies (e) and integrated intensities (f) as a function of $V_C$ with constant $V_G = 0V$ for B = 0T (black) and B = 7T (grey). Values are extracted from fits to the spectra. The dashed line in (e) marks the anticipated bare Stark shift of IX extrapolated for B = 7T. The double-arrow highlights the corresponding blue-shift due to the IX density increase. (g) IX energies as function of magnetic field at $V_G = 0V$ for different central gate voltages $V_C = 0V$ (diamonds), -0.2V (triangles), -0.4V (squares), -0.6V (circles). The dotted line marks the energy 1.560 eV, which is the maximum IX energy at 7T. (h) Calculated IX density $\Delta n_{IX}$ as a function of the magnetic field indicating a magnetic field dependent increase of the IX density. The data are extracted from the change in IX emission energy as function of magnetic field for $V_C = -0.6V$ and $V_G = 0V$.

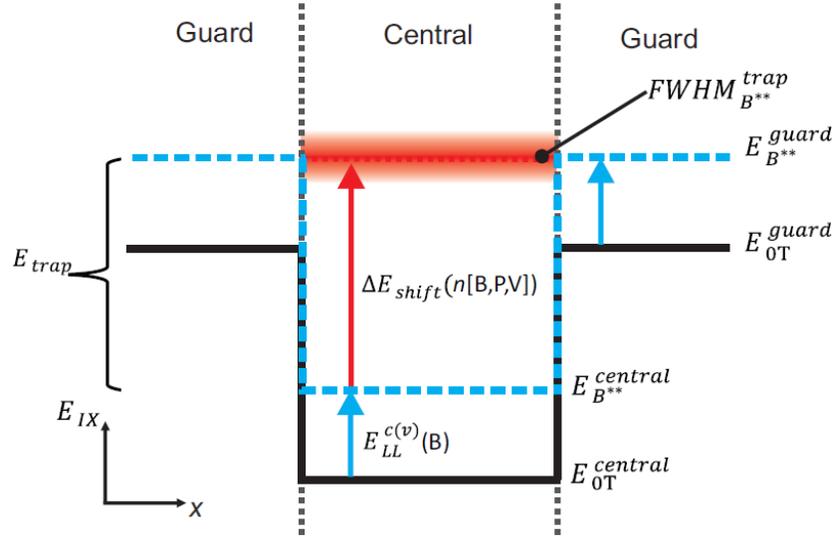

*Figure 3: (Color online) Schematic lateral IX potential landscape across the device for B = 0T (black solid line) and for B = 7T (dashed line). The ground state below the central (guard) gate electrode at B = 0T is indicated by the black line. The perpendicular magnetic field lifts the empty ground state by $E^{c(v)}_{LL}(B)$, due to LL formation independent from central and guard gate potential. The dipole-dipole interaction $\Delta E_{shift}$ between the IX increases with increasing IX density and results in a blue-shift of the IX emission energy. Since the IX intensity increases with increased applied magnetic field and consequently also $\Delta E_{shift}(n[P,V,B])$ increases until the depth of the trap potential ($E_{trap} = |V_G - V_C| \cdot ed/s$) is compensated for $E_{trap} \leq \Delta E_{shift}(B)$. In turn, the IXs do not feel any trap potential for magnetic field values larger than $B > B^{**}$. IX-IX scattering and background potential fluctuations cause a broadening of the energy levels contribute to the apparent energetic full width at half maximum in the emission ($FWHM^{trap}_{B^{**}}$).*

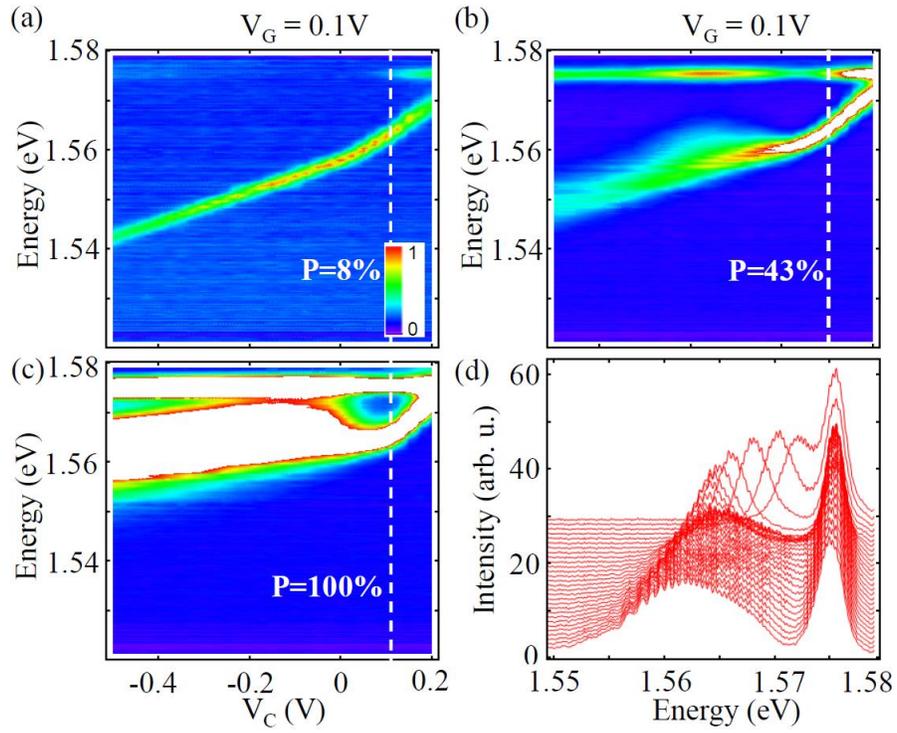

*Figure 4: (Color online) (a)-(c) Photoluminescence energy of IXs as function of $V_C$ for constant $V_G = 0.1V$ for different excitation powers (a: $P = 0.08 \cdot P_0$; b: $P = 0.43 \cdot P_0$; and c: $P = P_0$) at the critical magnetic field value $B^{**} = 7T$. Dashed vertical lines mark the transition between trap and anti-trap potential. (d) The same data as shown in (c) presented as a waterfall plot of single spectra.*